\documentclass[journal]{IEEEtran}

\usepackage[utf8]{inputenc}
\usepackage[T1]{fontenc}
\usepackage{microtype}
\usepackage{booktabs}
\usepackage{array}
\usepackage{xcolor}
\usepackage{listings}
\usepackage{tikz}
\usetikzlibrary{arrows.meta}
\definecolor{okblue}{RGB}{0,114,178}
\definecolor{okorange}{RGB}{230,159,0}
\definecolor{okverm}{RGB}{213,94,0}
\definecolor{okgreen}{RGB}{0,158,115}
\usepackage[hidelinks]{hyperref}
\usepackage{xurl}
\usepackage{needspace}

\newcommand{\code}[1]{\texttt{#1}}
\title{Spec Sheets Are Not Kernels: An ISA- and Source-Level Audit\\
of INT8 Availability on NVIDIA Blackwell Ultra}

\author{Teng-Ruei~Chen%
\thanks{T.-R. Chen is with Krixvon, Taipei 100, Taiwan (e-mail: luka@krixvon.com; ORCID: 0000-0001-8995-334X).}%
\thanks{Technical report, August 2026. All sources were accessed on 2026-08-11 (UTC) unless noted otherwise; repository findings are pinned to the exact commits and container digests listed in Appendix~\ref{app:repro}. This report deliberately contains \emph{no} performance measurements; a companion measurement study is in preparation.}}

\begin{document}
\maketitle
\bstctlcite{bstctl:audit}

\begin{abstract}
NVIDIA's published specifications give the Blackwell Ultra GPU (B300) a dense-compute ratio of roughly 30:1 between FP8 and INT8 tensor-core throughput; its predecessors, H200 and B200, both provide 1:1. We audit what this deprioritization means in practice by tracing INT8 W8A8 support through four layers of the stack: the published specifications, the PTX ISA, NVIDIA's CUTLASS kernel library, and the two major open-source LLM serving engines (vLLM and SGLang). We find a consistent, layered withdrawal: (i) the PTX ISA never exposes the fifth-generation tensor-core integer path (\code{tcgen05.mma} with \code{.kind::i8}) on \code{sm\_103a}, even though the same PTX revision extends the FP4 kinds to that target, leaving legacy warp-level IMMA as the only architecturally legal integer tensor-core path on B300; (ii) CUTLASS's kernel generator explicitly skips INT8 UMMA generation for any build targeting \code{103a}, while generating FP8 unconditionally; (iii) vLLM ships no INT8 GEMM for Blackwell and fails with a hard runtime error at the first forward pass, after the model has loaded; and (iv) SGLang's ahead-of-time INT8 GEMM stops at \code{Sm90}, while its FP8 tuning configurations already cover B200. We document an escape hatch (rerouting vLLM's INT8 path to a JIT-compiled Triton backend via an environment variable), a false-negative trap in the obvious profiler methodology for detecting ``native INT8'' on \code{sm\_103}, and the practical failure semantics that make naive testing expensive. Together, these findings show that a quantization format's availability is a property of the whole stack rather than of the model or the spec sheet. Four distinct layers, three of them NVIDIA's own, withdrew INT8 support in mutually consistent ways, and a format that is nominally present on the datasheet is, by default, undeployable on this hardware.
\end{abstract}

\section{Introduction}
\label{sec:intro}

Consider a practitioner who serves an LLM quantized to W8A8 INT8 on an H200 node and plans a migration to Blackwell Ultra (B300). Weights and activations are one byte per element in both INT8 and FP8~\cite{micikevicius2022fp8}, so memory footprint and memory-bound behavior carry over; the spec sheet lists an INT8 entry for the new GPU; the serving engine's documentation lists \code{w8a8\_int8} as a supported quantization. Every individual signal suggests the checkpoint will run.

In practice, the checkpoint does not run by default, and no fifth-generation tensor-core path exists for it. This report documents why, by auditing INT8 support layer by layer, from NVIDIA's published specifications down to the serving engines' kernel dispatch. The audit is entirely documentary: every claim below traces to an official NVIDIA document, to source code at a pinned commit, or to the contents of a shipped release binary, with URLs, commits, digests, and access dates collected in Appendix~\ref{app:repro}. We deliberately report \emph{no} performance measurements here.

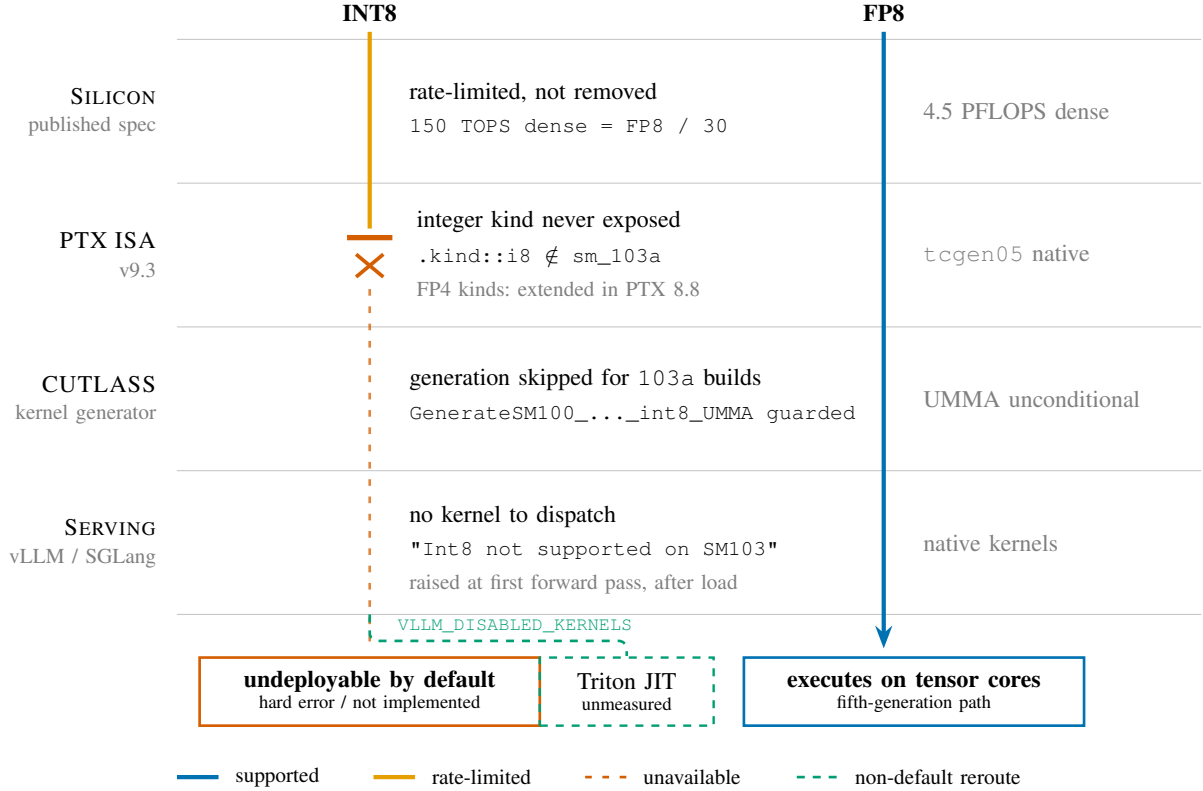
\begin{figure*}[t]
\centering
\begin{tikzpicture}[every node/.style={font=\small}]
  \foreach \y in {0,-1.9,-3.8,-5.7,-7.6} \draw[gray!45, thin] (2.55,\y) -- (16.1,\y);
  \node[anchor=east] at (2.4,-0.75) {\textsc{Silicon}};
  \node[anchor=east, gray, font=\footnotesize] at (2.4,-1.15) {published spec};
  \node[anchor=east] at (2.4,-2.65) {\textsc{PTX ISA}};
  \node[anchor=east, gray, font=\footnotesize] at (2.4,-3.05) {v9.3};
  \node[anchor=east] at (2.4,-4.55) {\textsc{CUTLASS}};
  \node[anchor=east, gray, font=\footnotesize] at (2.4,-4.95) {kernel generator};
  \node[anchor=east] at (2.4,-6.45) {\textsc{Serving}};
  \node[anchor=east, gray, font=\footnotesize] at (2.4,-6.85) {vLLM / SGLang};
  \node[font=\small\bfseries] at (5.1,0.35) {INT8};
  \node[font=\small\bfseries] at (11.9,0.35) {FP8};
  \draw[okblue, line width=1.6pt, -{Stealth[length=3mm]}] (11.9,0.1) -- (11.9,-8.05);
  \node[anchor=west, gray] at (12.3,-0.95) {4.5 PFLOPS dense};
  \node[anchor=west, gray] at (12.3,-2.85) {\code{tcgen05} native};
  \node[anchor=west, gray] at (12.3,-4.75) {UMMA unconditional};
  \node[anchor=west, gray] at (12.3,-6.65) {native kernels};
  \draw[okorange, line width=1.6pt] (5.1,0.1) -- (5.1,-2.5);
  \draw[okverm, line width=2pt] (4.8,-2.62) -- (5.4,-2.62);
  \draw[okverm, line width=1.4pt] (4.92,-2.84) -- (5.28,-3.14);
  \draw[okverm, line width=1.4pt] (5.28,-2.84) -- (4.92,-3.14);
  \draw[okverm, line width=1pt, dash pattern=on 2.5pt off 4pt, opacity=0.75] (5.1,-3.3) -- (5.1,-8.05);
  \node[anchor=west] at (5.5,-0.7) {rate-limited, not removed};
  \node[anchor=west, font=\footnotesize\ttfamily] at (5.5,-1.15) {150 TOPS dense = FP8 / 30};
  \node[anchor=west] at (5.6,-2.4) {integer kind never exposed};
  \node[anchor=west, font=\footnotesize\ttfamily] at (5.6,-2.85) {.kind::i8 $\notin$ sm\_103a};
  \node[anchor=west, gray, font=\footnotesize] at (5.6,-3.3) {FP4 kinds: extended in PTX 8.8};
  \node[anchor=west] at (5.5,-4.5) {generation skipped for \code{103a} builds};
  \node[anchor=west, font=\footnotesize\ttfamily] at (5.5,-4.95) {GenerateSM100\_...\_int8\_UMMA guarded};
  \node[anchor=west] at (5.5,-6.3) {no kernel to dispatch};
  \node[anchor=west, font=\footnotesize\ttfamily] at (5.5,-6.75) {"Int8 not supported on SM103"};
  \node[anchor=west, gray, font=\footnotesize] at (5.5,-7.2) {raised at first forward pass, after load};
  \draw[okgreen, dashed, line width=1pt, rounded corners=3pt] (5.1,-7.6) -- (5.1,-7.95) -- (8.5,-7.95) -- (8.5,-8.15);
  \node[anchor=west, okgreen, font=\scriptsize\ttfamily] at (5.35,-7.75) {VLLM\_DISABLED\_KERNELS};
  \node[draw=okverm, line width=1pt, align=center, minimum width=4.5cm, minimum height=0.9cm] at (5.1,-8.62)
    {\textbf{undeployable by default}\\[-2pt]{\scriptsize hard error / not implemented}};
  \node[draw=okgreen, dashed, line width=0.9pt, align=center, minimum width=2.3cm, minimum height=0.9cm] at (8.5,-8.62)
    {Triton JIT\\[-2pt]{\scriptsize unmeasured}};
  \node[draw=okblue, line width=1pt, align=center, minimum width=4.5cm, minimum height=0.9cm] at (12.3,-8.62)
    {\textbf{executes on tensor cores}\\[-2pt]{\scriptsize fifth-generation path}};
  \draw[okblue, line width=1.6pt] (2.55,-9.75) -- (3.1,-9.75);
  \node[anchor=west, font=\footnotesize] at (3.2,-9.75) {supported};
  \draw[okorange, line width=1.6pt] (5.15,-9.75) -- (5.7,-9.75);
  \node[anchor=west, font=\footnotesize] at (5.8,-9.75) {rate-limited};
  \draw[okverm, line width=1pt, dash pattern=on 2.5pt off 4pt] (7.95,-9.75) -- (8.5,-9.75);
  \node[anchor=west, font=\footnotesize] at (8.6,-9.75) {unavailable};
  \draw[okgreen, dashed, line width=1pt] (10.75,-9.75) -- (11.3,-9.75);
  \node[anchor=west, font=\footnotesize] at (11.4,-9.75) {non-default reroute};
\end{tikzpicture}
\caption{The path of a W8A8 checkpoint through the B300 stack. FP8 traverses all four layers natively. INT8 is rate-limited in silicon (Section~\ref{sec:specs}); the terminator marks the ISA layer, where the fifth-generation integer path was never exposed (Section~\ref{sec:isa}); CUTLASS then generates no INT8 kernels for \code{103a} builds (Section~\ref{sec:cutlass}), and the serving engines have no kernel to dispatch (Section~\ref{sec:serving}). The dashed reroute is the non-default Triton JIT path of Section~\ref{sec:vllm}, whose attained behavior is deliberately left unmeasured here. Colors follow the Okabe--Ito palette; every state is also encoded redundantly by line style.}
\label{fig:stack}
\end{figure*}

Figure~\ref{fig:stack} summarizes the findings: the silicon rate-limits INT8 to $\sim$1/30 of FP8 (Section~\ref{sec:specs}); the PTX ISA never exposes the new tensor-core integer path on B300's architecture target, while extending FP4 support to it in the same revision (Section~\ref{sec:isa}); NVIDIA's own kernel library skips INT8 kernel generation for B300 builds (Section~\ref{sec:cutlass}); and the two major open serving engines ship no Blackwell-capable ahead-of-time INT8 GEMM at all; one errors only at the first forward pass, and the other raises a generic not-implemented error above Hopper (Section~\ref{sec:serving}). Section~\ref{sec:measurement} derives methodological consequences for anyone attempting to measure ``native INT8'' on this hardware, including a profiler criterion that avoids a false negative we believe is otherwise nearly guaranteed. Section~\ref{sec:discussion} discusses implications and states explicitly what this report does \emph{not} claim.

\paragraph{Why this is worth a report.}
A large public inventory of INT8 W8A8 checkpoints and calibration pipelines~\cite{llmint8,smoothquant,zeroquant}, and a growing family of phase-aware mixed-precision serving policies~\cite{pmpd,hmaserve}, implicitly assume that the precision formats they assign have execution paths on the deployment target. On Blackwell Ultra that assumption fails for INT8 in a way that is visible in none of the places practitioners usually look (model cards, serving-engine feature tables, spec-sheet presence), and visible in all of the places they usually do not (ISA target lists, kernel-generator gates, dispatch tables). Independently, to our knowledge no third-party measurement of B300 INT8 throughput exists. B300 has been measured for other quantities~\cite{yin2026serialized,kim2026b300field}, and the Blackwell microbenchmark studies we are aware of cover B200, H200, and consumer Blackwell without mentioning B300~\cite{jarmusch2025microbench,jarmusch2025blackwelldissect}. The documentary record below is therefore, to our knowledge, the only public evidence base on this specific question, and it has a shelf life, since any of these code paths may change. Pinning the state of the stack, with timestamps, is the contribution.

\section{What the published specifications say}
\label{sec:specs}

Table~\ref{tab:specs} collects NVIDIA's published per-GPU tensor-core peaks, converted to a single basis: \emph{dense, per GPU, HGX (air-cooled) SKU bin}. Basis discipline matters more than usual here, because the INT8 figures differ by SKU bin and are published in sparse form: the Blackwell Ultra datasheet lists \code{INT8 Tensor Core = 330 TOPS | 307 TOPS} for the GB300~NVL72 and HGX~B300 columns respectively, with the footnote \emph{``Specification in sparse. Dense is \textonehalf{} sparse spec shown.''}~\cite{ultradatasheet} The Blackwell Architecture Technical Brief states the same quantities directly in dense/sparse pairs~\cite{blackwellbrief}.

\begin{table*}[t]
\centering
\small
\caption{Published per-GPU \emph{dense} tensor-core peaks (HGX SKU bin held throughout). Sources: NVIDIA H200 product page~\cite{h200page}; NVIDIA Blackwell Architecture Technical Brief, Table~3~\cite{blackwellbrief}; Blackwell Ultra Datasheet~\cite{ultradatasheet}. All accessed 2026-08-11.}
\label{tab:specs}
\begin{tabular}{lcccc}
\toprule
GPU & FP8 dense & INT8 dense & FP8:INT8 & FP64 (tensor) \\
\midrule
H200 SXM & 1{,}979 TFLOPS & 1{,}979 TOPS\textsuperscript{$\dagger$} & \textbf{1:1} & 67 TFLOPS \\
HGX B200 & 4.5 PFLOPS & 4.5 POPS & \textbf{1:1} & 37 TFLOPS \\
HGX B300 & 4.5 PFLOPS & 0.15 POPS (150 TOPS) & \textbf{$\sim$30:1} & 1.2 TFLOPS \\
\bottomrule
\end{tabular}

\smallskip
{\footnotesize \textsuperscript{$\dagger$}NVIDIA's H200 page prints the INT8 row's unit as ``TFLOPS'' (sic); the conventional unit for integer operations, used here, is TOPS.}
\end{table*}

\begin{figure*}[t]
\centering
\begin{tikzpicture}[every node/.style={font=\small}]
  \foreach \v/\x in {0.1/2.9, 0.3/6.03, 1/9.45, 3/12.58, 10/16.0} {
    \draw[gray!30, thin] (\x,0.5) -- (\x,-2.75);
    \node[gray, font=\footnotesize] at (\x,-3.05) {\v};
  }
  \node[anchor=east] at (2.7,0) {H200 SXM};
  \node[anchor=east] at (2.7,-1.1) {HGX B200};
  \node[anchor=east] at (2.7,-2.2) {HGX B300};
  \draw[gray!60] (4.05,-2.2) -- (13.73,-2.2);
  \draw[{Stealth[length=2mm]}-{Stealth[length=2mm]}, gray] (4.35,-1.8) -- (13.4,-1.8);
  \node[gray] at (8.9,-1.5) {$\sim$30$\times$};
  \node[draw=okorange, line width=1.2pt, minimum size=8pt, inner sep=0pt] at (11.39,0) {};
  \node[draw=okorange, line width=1.2pt, minimum size=8pt, inner sep=0pt] at (13.73,-1.1) {};
  \node[draw=okorange, line width=1.2pt, minimum size=8pt, inner sep=0pt] at (4.05,-2.2) {};
  \fill[okblue] (11.39,0) circle (2.4pt);
  \fill[okblue] (13.73,-1.1) circle (2.4pt);
  \fill[okblue] (13.73,-2.2) circle (2.4pt);
  \node[anchor=west, gray, font=\footnotesize] at (11.75,0) {1:1 (markers coincide)};
  \node[anchor=west, gray, font=\footnotesize] at (14.1,-1.1) {1:1};
  \node[anchor=north, gray, font=\footnotesize] at (4.05,-2.5) {150 TOPS};
  \fill[okblue] (12.7,0.85) circle (2.4pt);
  \node[anchor=west, font=\footnotesize] at (12.9,0.85) {FP8};
  \node[draw=okorange, line width=1.2pt, minimum size=8pt, inner sep=0pt] at (14.35,0.85) {};
  \node[anchor=west, font=\footnotesize] at (14.55,0.85) {INT8};
  \node[gray] at (9.45,-3.6) {published dense per-GPU tensor-core peak (PFLOPS / POPS, log scale)};
\end{tikzpicture}
\caption{Published dense per-GPU tensor-core peaks (HGX SKU bin; values as in Table~\ref{tab:specs}). INT8 markers are open squares and FP8 markers are filled circles, so a 1:1 pair renders as a circle inside a square: FP8 and INT8 coincide on H200 and B200 and separate by $\sim$30$\times$ only on B300.}
\label{fig:specs}
\end{figure*}
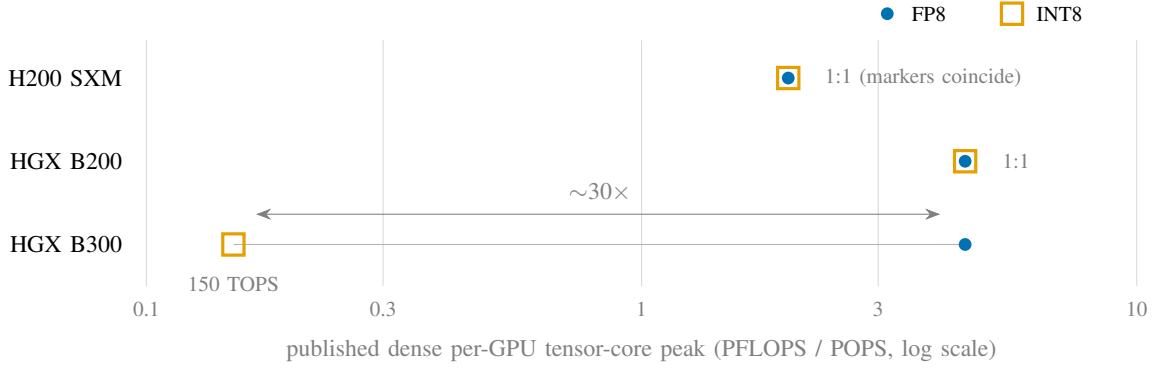

Three observations from Table~\ref{tab:specs} matter for the rest of the report.

\textbf{The discontinuity is B200$\to$B300, not Hopper$\to$Blackwell (Figure~\ref{fig:specs}).} H200 is 1:1 (the product page lists FP8 3{,}958 TFLOPS and INT8 3{,}958 ``TFLOPS'' (sic; NVIDIA's own label on the integer row), both footnoted \emph{``With sparsity.''}~\cite{h200page}); B200 is also 1:1 (Technical Brief Table~3: \code{FP8/FP6 Tensor Core Dense/Sparse 4.5/9 petaFLOPS}, \code{INT8 Tensor Core Dense/Sparse 4.5/9 petaOPS}~\cite{blackwellbrief}). The $\sim$30$\times$ cut is specific to Blackwell \emph{Ultra}. Any framing of the form ``Blackwell reduced INT8'' is therefore wrong on its face; we return to this in Section~\ref{sec:discussion}, because at least one published preprint propagates this error.

\textbf{FP64 was cut by essentially the same factor in the same product.} Technical Brief Table~3 lists FP64 tensor throughput falling from 37~TFLOPS (B200) to 1.2~TFLOPS (B300)~\cite{blackwellbrief}, and the CUDA Best Practices instruction-throughput table shows the SIMT-level counterpart (\code{add.f64}: 64 results/clock/SM at compute capability 10.0, 2 at 10.3)~\cite{cudabp}. INT8 and FP64 appear to have been deprioritized together, in favor of the FP4/FP8 inference envelope. This continues a longer integer-format retreat (INT4 tensor-core types were deprecated with Hopper), which industry analysis had already observed extending to Blackwell Ultra's INT8~\cite{semianalysis2025tensorcore}, and it matches NVIDIA's own push toward NVFP4 as its flagship low-precision format~\cite{nvidia2025nvfp4}. Nor was the direction inevitable: a dedicated hardware analysis had argued that INT8 is the more silicon-efficient of the two byte-wide inference formats~\cite{vanbaalen2023fp8int8}.

\textbf{INT8 is rate-limited, not removed.} The CUDA Programming Guide's Table~33 (``Input Data Types Supported by Tensor Core Acceleration per Compute Capability'') still lists INT8 = \emph{Yes} at compute capability 10.3, while FP64 is blank there~\cite{cudaguide}. Some tensor-core integer capability exists on the silicon. The question the rest of this report answers is: \emph{through which instructions, and reachable by which software?}

Two caveats we carry through the report. First, an FP8-to-INT8 ratio compares TFLOPS to TOPS, i.e.\ a rate ratio between two different operation types rather than a like-for-like FLOP count. Second, these are rounded marketing peaks (the datasheet and the brief disagree slightly, 307 vs.\ 300 TOPS sparse), so we write $\sim$30$\times$, never 30.3$\times$. We also note, without imputing intent, that NVIDIA's own Blackwell Ultra architecture blog post carries per-format comparison rows for NVFP4 and FP8 only (no INT8 row, no FP64 row), while the datasheet and the technical brief do publish the INT8 figures~\cite{ultrablog}.

\section{The ISA layer: \texttt{tcgen05.mma} has no integer kind on \texttt{sm\_103a}}
\label{sec:isa}

Blackwell introduces a fifth generation of tensor-core instructions, exposed in PTX as the \code{tcgen05.*} family; the matrix-multiply instruction \code{tcgen05.mma} selects its input class via a \code{.kind} qualifier (\code{.kind::f16}, \code{.kind::tf32}, \code{.kind::f8f6f4}, \code{.kind::mxf4nvf4}, \code{.kind::i8}, \ldots)~\cite{ptxisa}. B300 corresponds to compute capability 10.3, PTX target \code{sm\_103a}.\footnote{The cc-10.3$\leftrightarrow$B300 correspondence is corroborated independently: every B300 offer on a public GPU marketplace API reports \code{compute\_cap 1030} (observed 2026-08-11), and CUTLASS's only SM103-specific kernel generator is named for Blackwell Ultra's signature format (Section~\ref{sec:cutlass}).}

The PTX ISA manual (v9.3) states the supported targets for the integer kind verbatim:

\begin{quote}\small
``Qualifier \code{.kind::i8} is supported on following architectures: \code{sm\_100a} \code{sm\_101a} (Renamed to \code{sm\_110a} from PTX ISA version 9.0) \code{sm\_110a}''~\cite{ptxisa}
\end{quote}

\code{sm\_103a} is absent. The family-specific support statement carves the integer kind out explicitly: the phrase \emph{``except \code{.kind::i8}''} appears four times in the \code{tcgen05.mma} sections, once per variant (\code{.mma}, \code{.mma.sp}, \code{.mma.ws}, \code{.mma.ws.sp}; in the \code{.mma.sp} instance as part of the longer exclusion ``except \code{.kind::i8}/\allowbreak\code{.kind::mxf4nvf4}/\allowbreak\code{.kind::mxf4}'')~\cite{ptxisa}. And the manual's version-history table provides a controlled contrast \emph{within the same instruction family} (Table~\ref{tab:ptx}): in PTX ISA 8.8, NVIDIA extended the FP4 kinds (explicitly for the sparse variant \code{tcgen05.mma.sp}; for dense \code{tcgen05.mma} the FP4 kinds reach family targets via the family-wide statement above, whose only carve-out is \code{.kind::i8}), the \code{tcgen05.shift} instruction, and a new K=96 shape to \code{sm\_103a}, but did not extend \code{.kind::i8} anywhere.

\begin{table*}[t]
\centering
\small
\caption{PTX ISA version history for \code{tcgen05} features, from the manual's release-history table~\cite{ptxisa}, with each row attributed to its instruction variant. PTX 8.8 is the revision that added \code{sm\_103a} support; the integer kind was not extended to it under any variant.}
\label{tab:ptx}
\footnotesize
\begin{tabular}{llll}
\toprule
Feature (instruction: entry) & Introduced & Extended (PTX 8.8) & PTX 9.0 \\
\midrule
\code{tcgen05.mma}: base variant & 8.6 \code{sm\_100a} & \code{sm\_100f} & \code{sm\_110f} \\
\code{tcgen05.mma}: Kind \code{.kind::i8} & 8.6 \code{sm\_100a} & --- & \code{sm\_110a} \\
\code{tcgen05.mma}: K shape value 96 & --- & \textbf{\code{sm\_103a}} & --- \\
\code{tcgen05.mma.sp}: \code{.kind::mxf4nvf4}, \code{.kind::mxf4} & 8.6 \code{sm\_100a} & \textbf{\code{sm\_103a}} & \code{sm\_110a} \\
\code{tcgen05.shift}: base variant & 8.6 \code{sm\_100a} & \textbf{\code{sm\_103a}} & \code{sm\_110a} \\
\bottomrule
\end{tabular}
\end{table*}

The consequence: \textbf{on B300, the fifth-generation tensor-core integer path does not exist at the ISA level.} The only architecturally legal integer tensor-core path on \code{sm\_103a} is the legacy warp-level IMMA family (\code{mma.sync} with \code{.s8}/\code{.u8} operands, e.g.\ \code{mma.m16n8k32}), which the manual documents as requiring \code{sm\_80} or higher with no carve-out for \code{sm\_103} that we could find~\cite{ptxisa}, consistent with the Programming Guide's ``INT8 = Yes at cc 10.3''~\cite{cudaguide}. (Bit-exact \emph{emulation} of INT8 GEMM on FP4 tensor cores has recently been demonstrated~\cite{hayashi2026fp4ozaki}; that is a software scheme above the ISA, not an integer MMA path.) This single fact explains, rather than merely describes, everything the software layers below do: there is no B300 instruction for them to target except the legacy path, on silicon that runs it at $\sim$1/30 the FP8 rate.

\section{The kernel-library layer: CUTLASS skips INT8 for \texttt{103a} builds}
\label{sec:cutlass}

CUTLASS (at commit \code{dcf215a}, 2026-08-06) does support fifth-generation integer MMA, \emph{on SM100}. Its Blackwell functionality documentation states that ``\code{tcgen05.mma} instructions support all legacy types (\code{tfloat32\_t}, \code{half\_t}, \code{bfloat16\_t}, \code{int8\_t}, \code{uint8\_t})'' and lists dense and sparse \code{int8\_t} GEMM rows with unit tests (\path{sm100_tensorop_gemm/s8_s8_void_s32.cu}; a sparse IMMA variant likewise)~\cite{cutlass}. SM100 and SM103 must therefore be kept distinct: ``Blackwell supports it'' is true for B200 and false for B300, and conflating the two produces the error discussed in Section~\ref{sec:discussion}.

For B300 targets the kernel generator is explicit. In \path{python/cutlass_library/generator.py}, \code{GenerateSM100()} guards INT8 UMMA generation behind the \emph{absence} of the family targets that include \code{103a}, while FP8 generation is unconditional:

\Needspace*{8\baselineskip}
\begin{lstlisting}
arch_family_cc = ['100f', '101f', '103a']
...
if not bool(set(manifest.compute_capabilities_feature_set).intersection(arch_family_cc)):
    GenerateSM100_TensorOp_int8_UMMA_gemm(manifest, cuda_version)
GenerateSM100_TensorOp_fp8_UMMA_gemm(manifest, cuda_version)
\end{lstlisting}

The sparse INT8 generator carries the same guard. (When building with CUDA~$\geq$~13, the source rewrites \code{101f} to \code{110f} in this list; \code{103a} is a member either way.) Moreover, the only SM103-specific generator in the entire file is
\begin{center}
\scriptsize\path{GenerateSM103_TensorOp_fp4_ultra_UMMA_gemm_with_block_scaled}
\end{center}
\noindent and the count of SM103 generators matching \code{int8|s8|i8} is zero~\cite{cutlass}. NVIDIA's own kernel library, in other words, implements the ISA reality of Section~\ref{sec:isa}: any build that targets B300 gets FP4 and FP8 tensor-core GEMMs, and no INT8 UMMA.

\section{The serving layer: vLLM and SGLang}
\label{sec:serving}

\subsection{vLLM: no Blackwell INT8 GEMM, and a hard error only at the first forward pass}
\label{sec:vllm}

We audited vLLM~\cite{vllm,kwon2023paged} at commit \code{6c95a641} (main, 2026-08-11) and cross-checked every finding against the shipped v0.27.1 release image (digest pinned in Appendix~\ref{app:repro}).

\textbf{Build gates.} vLLM's CUTLASS-based W8A8 GEMMs are arch-gated in \code{CMakeLists.txt}. The Hopper source list (\code{9.0a}) includes INT8 kernels (\path{scaled_mm_sm90_int8.cu}, \path{scaled_mm_azp_sm90_int8.cu}); the Blackwell SM100 and SM120 source lists contain FP8 sources only; and the legacy CUTLASS-2.x path is gated to \code{"7.5;8.0;8.7;8.9+PTX"} (Turing through Ada, excluding \code{sm\_100+}). No INT8 GEMM is compiled for any Blackwell target.

\textbf{Dispatch.} The SM100 dispatch file passes a null pointer for the integer case, with a comment saying so:

\Needspace*{5\baselineskip}
\begin{lstlisting}
dispatch_scaled_mm(c, a, b, a_scales, b_scales, bias,
  vllm::cutlass_scaled_mm_sm100_fp8,
  nullptr,  // int8 not supported on SM100
  vllm::cutlass_scaled_mm_blockwise_sm100_fp8);
\end{lstlisting}

\noindent
and the shared dispatch helper turns that null pointer into a hard runtime error: \emph{``Int8 not supported on SM\code{<}N\code{>}. Use FP8 quantization instead, or run on older arch (SM \code{<} 100).''} We verified that this error string, together with entry points for \code{sm75}--\code{sm120}, is present in the compiled extension (\path{_C_stable_libtorch.abi3.so}) inside the official release image; this is shipped behavior rather than a development-branch artifact. Note the comment's wording (``not supported on SM100'') describes vLLM's build configuration, not the hardware or CUTLASS: as Section~\ref{sec:cutlass} shows, SM100 INT8 UMMA exists upstream; vLLM does not instantiate it, and for SM103 there is nothing to instantiate.

\textbf{Failure semantics.} The failure arrives late. The INT8 W8A8 quantization scheme's Python-side gate, \path{get_min_capability()}, returns 75 (comment: ``turing and up''), and the first-priority kernel's \path{can_implement()} returns \code{(True, None)} unconditionally. A B300 (cc 10.3 $\geq$ 7.5) therefore passes every load-time check: the checkpoint downloads, weights load, and the process dies at the \emph{first forward pass}. For a 405B-class model this means paying for the full download and load before learning the deployment is impossible; we quantify the methodological consequence in Section~\ref{sec:measurement}.

\textbf{The escape hatch.} vLLM selects INT8 linear kernels from a priority list (the Cutlass, then Triton, then Humming variants of \code{Int8ScaledMMLinearKernel} on CUDA platforms), and its selector honors an environment variable, \code{VLLM\_DISABLED\_KERNELS}, that removes named kernels from consideration. Setting

\begin{lstlisting}
VLLM_DISABLED_KERNELS=CutlassInt8ScaledMMLinearKernel
\end{lstlisting}

\noindent
reroutes INT8 W8A8 to the Triton backend, whose kernels are JIT-compiled for the running architecture rather than AOT-gated; no code patch is required. We functionally verified the mechanism on an \code{sm\_89} (Ada) GPU using the release image: the three configurations (default; Cutlass disabled; Cutlass and Triton disabled) select the three respective backends (vLLM logs an audit line, \code{Selected <Kernel> for <Scheme>}, at INFO level), and all three produce identical greedy-decoded output on a fixed prompt. Whether the Triton path emits actual IMMA instructions \emph{on \code{sm\_103}}, and at what attained throughput, is the open measurement question this report leaves to the companion study.

\textbf{A structural confound for anyone comparing formats.} vLLM's FP8 kernel priority list for per-tensor/per-channel scaling (\code{Marlin}, \code{FlashInfer}, \code{Cutlass}, two Torch fallbacks, \code{Humming}) contains \emph{no Triton FP8 backend}; a Triton FP8 kernel exists only in the separate \emph{blockwise}-scaling list (\path{TritonFp8BlockScaledMMKernel}). On B300, therefore, an INT8-vs-FP8 comparison under matched per-tensor/per-channel recipes is \emph{by construction} also a Triton-vs-native-backend comparison (Marlin, FlashInfer, or CUTLASS, whichever the selector picks for FP8). A fully backend-matched comparison at matched scale granularity may simply be unavailable; measurement studies must either bound the backend effect (e.g., by measuring the Cutlass-vs-Triton INT8 delta on H200, where both exist) or report B300 INT8 results as path-specific bounds rather than format properties.

\subsection{SGLang: AOT INT8 stops at Hopper; FP8 tuning already covers B200}
\label{sec:sglang}

SGLang~\cite{sglang,zheng2023sglang} (at commit \code{b20c375}, 2026-08-11) ships no Blackwell-capable ahead-of-time INT8 GEMM. Its CUTLASS-based INT8 kernel (\code{int8\_gemm\_kernel.cu}; full path in Appendix~\ref{app:repro}) stops at Hopper: the file contains \code{cutlass::arch} tags \code{Sm75} ($\times$1) and \code{Sm80} ($\times$6), a single \code{cutlass::arch::Sm90} tag (plus ten \code{Sm90}-prefixed epilogue-type names), and none of \code{Sm100}/\code{Sm103}/\code{Sm120}. The newest architecture-specific dispatch entry points are the \code{sm90} ones (\path{cutlass_int8_scaled_mm_sm90}, \path{sm90_dispatch_*}); \code{sm75}/\code{sm80}/\code{sm89} variants cover older architectures, and nothing above \code{sm90} exists. On any newer architecture the public entry \code{int8\_scaled\_mm} falls through to a generic error,
\Needspace*{4\baselineskip}
\begin{lstlisting}
TORCH_CHECK_NOT_IMPLEMENTED(false, "No implemented int8_scaled_mm for current compute capability.");
\end{lstlisting}
\noindent without the prescriptive use-FP8 guidance vLLM emits. As with vLLM, an architecture-agnostic Triton JIT kernel does exist (a \emph{blockwise} INT8 GEMM, \path{w8a8_block_int8_matmul}); what does not exist is any Blackwell tuning configuration for it, which brings us to the natural control below.

The repository also offers a natural control on maintainer investment: the tuning-configuration files for these Triton kernels are named by device and dtype, and their coverage is asymmetric (Table~\ref{tab:sglang}). Same repository, same directory, same naming convention:
\begin{table}[t]
\centering
\footnotesize
\caption{Device coverage of SGLang's bundled kernel-tuning configurations, by dtype (commit \code{b20c375}; directory in Appendix~\ref{app:repro})~\cite{sglang}.}
\label{tab:sglang}
\begin{tabular}{@{}l>{\raggedright\arraybackslash}p{5.4cm}@{}}
\toprule
Tuning configs & Devices covered \\
\midrule
\code{int8\_w8a8} & A100-SXM4-80GB, A800-SXM4-80GB, H20 \\
\addlinespace[2pt]
\code{fp8} & \textbf{B200}, H100\_80GB\_HBM3, H200, H20, L20Y, L40, MI300X, MI325X, \ldots \\
\bottomrule
\end{tabular}
\end{table} FP8 tuning has been carried forward to Blackwell, while INT8 tuning stopped before Hopper's flagship parts. No one asserts this pattern anywhere; it falls out of routine maintenance, which is what makes it credible evidence of where the ecosystem's effort is going.

\section{Consequences for measurement studies}
\label{sec:measurement}

Three methodological consequences follow directly from the audit; we state them as guidance because we expect measurement work (ours included) to follow this report.

\textbf{(1) The obvious profiler criterion produces a false negative on \code{sm\_103}.}
A natural check for ``native INT8'' on Blackwell would be to profile the workload and grep kernel names or instruction mixes for the fifth-generation path (\code{tcgen05}, or \code{sm103}-specific kernels). Section~\ref{sec:isa} shows why this must fail: on \code{sm\_103a} there is no \code{tcgen05} integer kind for any kernel to use, so \emph{a genuinely native INT8 execution will show none}, and would be misclassified as a fallback. The reliable criterion is the IMMA instruction counter: in Nsight Compute, a nonzero \path{sm__inst_executed_pipe_tensor_op_imma.sum} (equivalently, IMMA-class SASS in the instruction mix), combined with int8 operand dtypes and the absence of dequantize-to-BF16 kernels between quantized layers. Kernel-name string matching is not a valid criterion on this architecture; instruction-level verification follows established tensor-core microbenchmarking practice~\cite{sun2022tensorcores}.

\textbf{(2) Late failure makes naive testing expensive; use a small fixture first.}
Because vLLM's INT8 path passes all load-time checks and fails at the first forward pass (Section~\ref{sec:vllm}), the cheap and correct smoke test on any Blackwell rental is a $\sim$1B-parameter INT8 W8A8 checkpoint, e.g.\ the publicly available \path{RedHatAI/Llama-3.2-1B-Instruct-quantized.w8a8} (channel-wise symmetric weights, dynamic per-token activations, \code{lm\_head} excluded; produced by the same team and W8A8 recipe evaluated in~\cite{kurtic2024bf16}). It exercises the entire failure path (download, load, first forward pass) at $\sim$1B scale, so the failure, where present, surfaces in minutes rather than after a multi-hundred-gigabyte download at node-hour prices.

\textbf{(3) Pin container digests; \code{:latest} drifts fast enough to change the experiment.}
During this audit, the \code{vllm/vllm-openai:latest} tag moved underneath us within hours, from a digest carrying vLLM 0.23.0 (with \emph{two} INT8 CUDA backends in the priority list) to one carrying 0.27.1 (with three). Any experiment that does not pin \code{image@sha256:\ldots} may silently change kernel-selection behavior between runs, or between the two nodes of a two-hardware comparison, where the difference would masquerade as a hardware effect. The digests we audited are listed in Appendix~\ref{app:repro}.

\section{Discussion}
\label{sec:discussion}

\textbf{Availability is a stack property.}
The spec sheet says B300 ``has'' INT8; the Programming Guide agrees; the serving engine's feature table lists \code{w8a8\_int8}. All are true, and yet the default result of deploying an INT8 W8A8 checkpoint on B300 is a runtime error after a complete model load. The gap opens because actors at four distinct layers (silicon architects, ISA designers, and kernel-library maintainers within NVIDIA, and serving-framework maintainers outside it) each made a locally rational decision given a 30:1 rate disadvantage, and the decisions compose into de facto unavailability. None of these decisions is visible at the level where deployment decisions are usually made. The episode is a small, unusually well-documented instance of the broader pattern in which hardware and software support, rather than intrinsic merit, decide which methods remain viable~\cite{hooker2020hardware}. In ML systems specifically, a monoculture of a few hyper-optimized kernels has likewise been observed to steer what is explorable at all~\cite{barham2019rut}. We suggest treating \emph{kernel coverage per (format, architecture)} as a first-class property in capacity planning, alongside the memory and peak-rate arithmetic that roofline-style planning models already capture~\cite{williams2009roofline,llmviewer,genz}; SGLang's tuning-config inventory (Section~\ref{sec:sglang}) shows this can be read directly off maintenance artifacts.

\textbf{The guardrail: this is a Blackwell-\emph{Ultra} property.}
Because H200 and B200 are both 1:1, nothing in this report supports a ``Hopper$\to$Blackwell'' generational narrative, and we caution against one explicitly because the error is already in circulation: a recent preprint arguing for FP8-centric HPC~\cite{matsuoka2026fp8} correctly reports the B300 cut ($\sim$165 TOPS INT8 vs.\ 5 PFLOPS FP8, GB300 bin) but tabulates B200 INT8 as $\sim$155 TOPS, low by roughly 29$\times$ against NVIDIA's published 4{,}500 TOPS dense~\cite{blackwellbrief} and against measured B200 results ($\approx$3{,}928 TOPS)~\cite{jarmusch2025microbench}. The INT8 asymmetry begins, and so far ends, at Blackwell Ultra. (Rubin-generation claims are outside our scope; the Ozaki-scheme literature observes, from NVIDIA's published specifications, that INT8 performance is reduced on Blackwell Ultra \emph{and Rubin}, making reliance on INT8 alone insufficient~\cite{uchino2026ozaki}. That assessment is consistent with, but not established by, anything we audit here; the newest installment of that literature already emulates INT8 GEMM bit-exactly on FP4 tensor cores~\cite{hayashi2026fp4ozaki}.)

\textbf{What this means for quantization research and practice.}
Phase-aware and mixed-precision serving policies assign different reduced-precision formats to different phases or modules~\cite{pmpd,hmaserve}; any such policy implicitly assumes its assigned formats execute on tensor cores wherever they are placed. Consider a policy that assigns INT8 W8A8, the format NVIDIA itself standardized for production inference~\cite{wu2020integer}, with roots in integer-arithmetic-only inference~\cite{jacob2017quantization} and descendants ranging from quantization--system co-designs~\cite{qserve} to deployed INT8 attention kernels whose fast path is IMMA~\cite{sageattention}. On B300, by default, that format executes nowhere. More broadly: INT8 and FP8 are byte-identical in storage, so \emph{capacity} planning transfers across the H200$\to$B300 migration while \emph{compute-path} planning silently does not. That asymmetry is invisible in any single document we audited; it is visible only in their conjunction.

\textbf{What we do not claim.}
(i) We report no throughput, latency, or accuracy measurements; in particular we do \emph{not} claim the $\sim$30:1 spec ratio is attained in practice, in either direction. To our knowledge no third-party B300 INT8 measurement exists (B300 has been measured for other quantities, e.g.\ BF16 matmul and NVLink throughput under confidential computing~\cite{yin2026serialized}; INT8 has not~\cite{jarmusch2025microbench}), the absence argues for performing the measurement; it proves nothing by itself. (ii) INT8 is not ``removed'' from B300: the legacy warp-level IMMA path is architecturally present~\cite{ptxisa,cudaguide}. (iii) NVIDIA publishes the relevant INT8 figures in its datasheet and technical brief~\cite{ultradatasheet,blackwellbrief}; nothing here supports a concealment narrative. (iv) TensorRT-LLM is absent from our source-level claims; we audited CUTLASS, vLLM, and SGLang only. (v) All repository findings are point-in-time, pinned in Appendix~\ref{app:repro}; the gaps documented here are the kind that a single merged pull request can change; that is part of why this report exists.

\textbf{Outlook.} A companion study, in the microbenchmarking tradition of GPU architecture dissection~\cite{jia2018volta,luo2024hopper,jarmusch2025microbench}, will measure what this report only delimits: attained INT8 throughput on B300 through the two remaining paths (warp-level IMMA microbenchmarks; the Triton reroute of Section~\ref{sec:vllm}), against H200 and against FP8/NVFP4, where promise-versus-attained FP4 gaps are already documented on B200~\cite{egiazarian2025fp4gap}, under matched quantization recipes.

\appendix[Reproducibility: sources, pins, and verification commands]
\label{app:repro}

All URLs accessed 2026-08-11 (UTC) unless noted. The verification scripts, the profiler criterion of Section~\ref{sec:measurement}, and the pre-registered protocol of the companion measurement study are public at \url{https://github.com/luka-krixvon/unequal-tensor-cores}. Items marked \emph{author-local} (the contents of pulled container images, the \code{sm\_89} functional run, and the marketplace snapshot) are corroborated by pinned public sources where applicable but are not independently re-checkable from public records.

\subsection*{Official documents}
\begin{itemize}\setlength\itemsep{0.1em}
  \item PTX ISA v9.3: \url{https://docs.nvidia.com/cuda/parallel-thread-execution/index.html}. Key strings: ``Qualifier .kind::i8 is supported on following architectures:\ldots'' (integer-kind target list, \code{sm\_103a} absent); ``except .kind::i8'' ($\times$4, one instance embedded in a longer exclusion); the \code{tcgen05} release-history table rows excerpted in Table~\ref{tab:ptx}.
  \item Blackwell Architecture Technical Brief (PDF via \url{https://resources.nvidia.com/en-us-blackwell-architecture/blackwell-architecture-technical-brief}), Table 3, rows labeled \emph{Dense/Sparse}: HGX B300 FP8/FP6 4.5/9 petaFLOPS; INT8 0.15/0.30 petaOPS; FP64 1.2 teraFLOPS; HGX B200 4.5/9 petaFLOPS; 4.5/9 petaOPS; 37 teraFLOPS.
  \item Blackwell Ultra Datasheet (via \url{https://resources.nvidia.com/en-us-blackwell-architecture/blackwell-ultra-datasheet}), p.~4: INT8 330 / 307 TOPS (GB300 / HGX B300), footnote ``Specification in sparse. Dense is \textonehalf{} sparse spec shown.''
  \item H200 product page: \url{https://www.nvidia.com/en-us/data-center/h200/} (SXM column; footnote ``With sparsity.''; the INT8 row's unit is printed ``TFLOPS'', sic).
  \item CUDA Programming Guide, compute-capability appendix (Table 33): \url{https://docs.nvidia.com/cuda/cuda-programming-guide/05-appendices/compute-capabilities.html}.
  \item CUDA Best Practices Guide, arithmetic-instruction throughput (Table 5): \url{https://docs.nvidia.com/cuda/cuda-c-best-practices-guide/}. (SIMT instructions only; carries no tensor-core rows; do not cite it for matrix rates.)
  \item NVIDIA developer blog, ``Inside NVIDIA Blackwell Ultra'': \url{https://developer.nvidia.com/blog/inside-nvidia-blackwell-ultra-the-chip-powering-the-ai-factory-era/} (comparison table has NVFP4 and FP8 rows; no INT8 or FP64 rows).
\end{itemize}

\subsection*{Repositories and binaries (pinned)}
\begin{itemize}\setlength\itemsep{0.1em}
  \item CUTLASS \code{dcf215a} (2026-08-06): \path{media/docs/cpp/blackwell_functionality.md} (SM100 int8 rows and unit tests); \path{python/cutlass_library/generator.py}, \code{GenerateSM100()} (INT8-UMMA guard on \code{\{100f,101f,103a\}}, with \code{101f} rewritten to \code{110f} when building with CUDA~$\geq$~13 (\code{103a} is in the set either way); FP8 unconditional; sole SM103 generator is FP4).
  \item vLLM \code{6c95a641} (2026-08-11; full hash below):\\[0.2em]
  {\footnotesize\texttt{6c95a641e95c0faa6f3aa802d1fd3cce3f3bc3ce}}\\[0.2em] \code{CMakeLists.txt} (SM90 int8 sources; SM100/SM120 fp8-only; c2x gate \code{"7.5;8.0;8.7;8.9+PTX"}); \path{csrc/libtorch_stable/quantization/w8a8/cutlass/scaled_mm_c3x_sm100.cu} (\code{nullptr} int8 handler); \path{csrc/.../c3x/scaled_mm_helper.hpp} (error string); \path{vllm/model_executor/layers/quantization/compressed_tensors/schemes/compressed_tensors_w8a8_int8.py} (\code{get\_min\_capability()=75}); \path{vllm/model_executor/kernels/linear/__init__.py} (INT8/FP8 priority lists; \code{VLLM\_DISABLED\_KERNELS} handling).
  \item vLLM release image (v0.27.1, torch 2.13.0+cu130; digest public, binary contents \emph{author-local}):\\[0.2em]
  {\scriptsize\texttt{vllm/vllm-openai@sha256:}\\
  \texttt{0a51ea5b4ae2dc5d81890e5173f54203}\\
  \texttt{d2a3ae0cfffe51b8fd2afd4391bfd967}}\\[0.2em]
  \code{strings \_C\_stable\_libtorch.abi3.so} contains ``Int8 not supported on SM''; entry points present for \code{sm75}/\code{sm80}/\code{sm89}/\code{sm90}/\code{sm100}/\code{sm120} (\code{cutlass\_scaled\_mm\_sm*}). Earlier the same day, \code{:latest} resolved to \code{sha256:6d8429e3\ldots} (v0.23.0, torch 2.11.0+cu130, two INT8 backends); this is the drift noted in Section~\ref{sec:measurement}.
  \item SGLang \code{b20c375} (2026-08-11): \path{python/sglang/kernels/aot/csrc/gemm/int8_gemm_kernel.cu} (arch-tag census); \path{python/sglang/kernels/ops/quantization/configs/} (device-named tuning files; \code{int8\_w8a8} vs.\ \code{fp8} coverage sets as in Section~\ref{sec:sglang}).
  \item Functional reroute check (\emph{author-local}; no performance claims): \code{sm\_89} GPU, release image above, model \path{RedHatAI/Llama-3.2-1B-Instruct-quantized.w8a8}; three runs (default / Cutlass disabled / Cutlass+Triton disabled) select Cutlass, Triton, Humming INT8 kernels respectively (log line \code{Selected <Kernel> for CompressedTensorsW8A8Int8}) and produce identical greedy output. The selection mechanism and log line are independently verifiable in source at the pinned commit.
  \item B300 = cc 10.3 corroboration (\emph{author-local} marketplace snapshot): all B300 offers on the Vast.ai public bundles API reported \code{compute\_cap 1030} and \code{gpu\_ram 275040}~MiB ($\approx$268.6 GiB) on 2026-08-11.
\end{itemize}

\bibliographystyle{IEEEtran}
\bibliography{refs}

\end{document}